**BETWEEN A ROCK AND HARD PLACE: ASSESSING THE APPLICATION OF DOMESTIC POLICY AND SOUTH AFRICA'S COMMITMENTS UNDER THE WTO'S BASIC TELECOMMUNICATIONS AGREEMENT**

*Tracy Cohen∗*

**ABSTRACT**

*South Africa adopted the GATS Basic Agreement on Telecommunications and the regulatory principles in 1998. Obligations undertaken by South Africa mirrored the framework for the gradual telecommunications reform process that was begun in 1996. In the light of two threatened actions for anti-competitive practices in violation of the Agreement, this paper reviews the nature of the commitments undertaken by South Africa and assesses the country's compliance to date. This paper also seeks to explore the tension that arises between domestic policy reforms and international trade aspirations. It is argued that the dynamic produced through this tension affords domestic governments a mechanism with which to balance the seemingly opposing goals of competition and development. It is further argued that the broad regulatory principles, adopted by all signatories and often criticized for lack of precision, facilitate this fine balancing and affords domestic governments an opportunity to advance sovereign concerns while pursuing international trade ideals.*

## I. INTRODUCTION

More commonly known as the Basic Telecommunication Agreement or GATS[1] Fourth Protocol,[2] this agreement is the product of efforts by members of the World Trade Organization (WTO) to introduce global competition in basic telecommunications services. Three years later, this paper reviews the extent to which South Africa as a signatory since February 1998, has complied with that undertaking. It is suggested that while SA may not have reached full compliance, the tension produced by competing domestic policy goals on the one hand, and international trade aspirations on the other, accounts for this. I will argue however, that it is not necessary to resolve this tension as neither the domestic or international policy frameworks on their own will produce the goals and ideals espoused by GATS or domestic government's liberalization policies.[3] Rather, it is asserted that in the nexus of domestic development and international trade lies a useful mechanism of checks and balances. If carefully crafted and astutely applied this can facilitate broader sovereign domestic policy whilst honouring commitments to the multilateral trading order.

∗ Tracy Cohen, Graduate Fellow and Doctoral Candidate, Centre for Innovation Law and Policy, Faculty of Law, University of Toronto. Email: t.cohen@utoronto.ca. The author thanks Hudson Janisch and Michael Trebilcock for their valuable input and comment; Yasmin Carrim, Kobus Du Plooy, Alison Gillwald and Myron Zlotnick for assistance with current documents and developments. Any errors are those of the author.

[1] *General Agreement on Trade in Services,* contained as Annex 1(b) to *General Agreement on Tariffs and Trade – Multilateral Trade Negotiations (The Uruguay Round): Final Act Embodying the Results of the Uruguay Round of Trade Negotiations,* 15 December 1994, 33 I.L.M. 1167, 1195.
[2] *Fourth Protocol to the Basic Agreement on Trade in Services,* 30 April 1996, WTO Doc. S/L/20.

Part I of this paper sketches the main tenets of the GATS and its related instruments relevant to telecommunications. A brief outline of telecommunications reform in SA relevant to an understanding of the implications for international trade in also presented. Part II will enumerate SA's commitments under the Fourth Protocol and the Schedule of Specific Commitments, (the Schedule) including the additional obligations undertaken by adoption of the Regulatory Reference Paper (RP). A comprehensive analysis of these undertakings will be provided in regard to the current status of each within SA. While progress has been made on all six principles in the RP, it is submitted that a weak regulatory agency, beset with adverse internal and external influences has had the most adverse impact on the pace of change. The competitive safeguard dispute that arose between the United States and SA in respect of Telkom SA Ltd and AT&T Global Networks is instructive and will be reviewed.

In Part III, the "global-local" matrix is explored practically with reference to the conflicting domestic goals contained in the objectives of the 1996 South African Telecommunications Act.[4] The perceived limitations to the Fourth Protocol and the RP in particular are also examined. It is argued that acknowledging the distinctive nature of trade in services contributes to an understanding of the tensions that emerge. This in turn necessitates flexibility for the domestic application of global trade ideals in order to cater for divergent social and economic priorities.[5] It is however conceded that that while the effect of the tension between global and local policies can be beneficial, similar tensions produced by conflicting national goals on the domestic level is of less value. To the extent that this tension affects global participation in the SA telecommunications market, its speedy resolution should be a priority.

---

[3] Including competition, attracting foreign investment or increasing network rollout.
[4] The Telecommunications Act, 1996.
[5] On the broader sovereignty debate, see generally John O. McGinnis, 'The Political Economy of Global Multilateralism', 1 Chi J Intl Law 2  at 381 (2000) and Kal Raustiala, "Sovereignty and Multilateralism" 1 Chi J Intl Law 2 at 401 (2000).

## II. OVERVIEW

Enormous technological developments and regulatory reforms over the last decade have supported a paradigm shift that has placed trade in services firmly within the WTO agenda.[6] Consensus on the framework for such trade has evolved over a number of years and is reflected in a set of interrelated Agreements and Annexes: GATS; the Annex on Telecommunications (the Annex); the Fourth Protocol on Basic Communications; and the Reference Paper on Regulatory Principles. (RP)[7]

## A. GATS and the Annex on Telecommunications[8]

The GATS covers all services[9] and defines them by reference to the four modes of supply characteristic of service industries: cross border, consumption abroad, commercial presence in the consuming country and the presence of natural persons. Distinguishing these modes is crucial to recognizing that different forms of trade carry distinct domestic implications and create specific regulatory concerns across borders. These issues have shaped both the principles and rules embodied in the GATS, as well as the specific commitments that WTO members have undertaken in their Schedules.[10]

---

[6] Services were historically viewed as non-tradable for a range of legal, institutional, economic and technical reasons including their supposed intangibility or invisibility; their non-durable or transitory character and the view that they require direct and simultaneous production and consumption. See E. Senunas, 'The 1997 GATS Agreement on Basic Telecommunications: A Triumph for Multilateralism, or the Market' B.C. Intell. Prop. & Tech. F. 111, 401 (1997). For the arguments excepting to this characterization, see Michael J. Trebilcock and Robert Howse, *The Regulation of International Trade* (2d edn, London: Routledge, 2000) at 271-73. See also WTO Secretariat, 'An Introduction to the GATS', World Trade Organization, available at http://www.wto.org/english/tratop_e/serv_e/gsintr_e.doc (visited 3 November 2000).

[7] *Reference Paper on Regulatory Principles used for consideration as additional commitments in offers on basic telecommunications,* WTO Negotiating Group on Basic Telecommunications (24 April 1996). See http://www.wto.org/wto/press/refpap-e.htm. (last modified: 25 April 1997).

[8] *Annex on Telecommunications,* April 1994, App. To GATS, at http://wto.org/services/12-tel.htm (last modified: 25 April 1997). The current membership of the WTO is 140, as of 30 November 2000, of which over 100 are developing countries or countries in transition. For an overview of the GATS and trade in services, see generally Trebilcock and Howse, above note 6 at 270ff. See also Michael. H. Ryan 'Trade in Telecommunications Services: A Guide to the GATS' (Coudert Brothers: London, 1997).

[9] Except those that are "supplied in the exercise of governmental authority which are neither supplied on a commercial basis nor in competition with other service suppliers." GATS, Article I:1 and I:3.

[10] GATS can be simplified into two major components: the framework agreement containing the general obligations and disciplines that apply to all WTO Members (Part II comprising Articles II - XV inclusive) and the specific

GATS is modeled on the GATT and is similarly imbued with the canon of non-discrimination through the principles of "most-favoured nation" (MFN) and National Treatment.[11] Respectively, this requires that a concession extended by one Member to any other must immediately and unconditionally be extended to all other Members,[12] and that foreign services and suppliers be treated no less favorably than domestic ones.[13] Access to foreign markets is supported by a specific commitment requiring all Members to refrain from imposing any quantitative restrictions, economic needs test, or local incorporation requirements on another Member unless such limitations have specifically been listed in that Member's Schedule.[14]

A number of general obligations and disciplines in the GATS and the Annex are also important to telecommunications apart from and in addition to the Schedules. These include requirements that all laws affecting service provision be publicly available[15] and that domestic regulation be reasonable, objective, impartial and not more burdensome than necessary.[16] In suggesting a standard of proportionality in this regard, the GATS explicitly acknowledges that the pursuit of national interests will vary across borders and require different approaches, oftentimes discordant with the GATS ideals. Proportionality, or what constitutes "burdensome" regulation, allows an objective gauge to be used in assessing acceptable limits in this regard.[17]

While telecommunications services in most countries remains either fully or state owned, Article VIII and IX prohibit an incumbent from generally abusing its monopoly position or engaging in any anti-competitive business practices, such as price collusion or vertical integration in service sectors where that incumbent competes with other non-exclusive service

---

commitments that are identified in the individual WTO Member Schedules (Part III encompassing Articles XVI-XVII inclusive.)

[11] Unlike the GATT, GATS allows for differential application of the two: MFN applies horizontally while National Treatment is undertaken as a specific commitment.

[12] The 'unconditionality' principle can however, be narrowed through exemption, Article II:2. See also Ryan, above note 8, and Arlan Gates, 'Technological Change, Industry Concentration and Competition Policy in the Telecommunications Sector' 58 Univ. Toronto Faculty of L. R. 2 at 98 (2000).

[13] Article XVII:1.

[14] Article XVI: 2 (a)-(f). General exceptions exist, including measures necessary to protect public morals and order, protect human and animal health or secure compliance with non-discriminatory laws and regulations; national security or to secure obligations under the UN Charter to maintain international peace and security. The specific commitments, including national treatment, apply only to the scheduled or listed sectors and thus only bind members to the extent that they undertake commitments.

[15] GATS Article III:3 and the Annex, provision 4.

[16] Article VI:1and 4.

[17] Janisch notes the value however of the proportionality rule in keeping disproportionate licensing requirements in check. See Hudson N. Janisch, 'International Influences on Communications Policy in Canada' in Orr & Wilson (eds), *The Electronic Village, Policy Issues of the Information Economy* (Toronto: C.D. Howe Institute, 1998) at 76.

suppliers. In the SA context, the existence of Telkom's business unit, SAIX and its subsidiary Intekom, in the Internet industry, has raised questions regarding the extent to which their exclusivity is used to leverage advantage in the highly competitive Internet service provision (ISP) market.[18] These provisions will be considered in greater detail below, as they form the kernel of the threatened complaint to the WTO by the US Trade Representative against SA, on behalf of AT&T. They also warrant consideration because in the context of gradual liberalization, the grace period afforded to Telkom in which to rebalance tariffs towards costs,[19] coupled with the lack of public access to accounting records, renders the Article VIII and IX provisions of the GATS particularly difficult to monitor.

## B. The Fourth Protocol and the Regulatory Reference Paper

At the close of the Uruguay Round in 1994, gains had only been made in the supply of value-added services, with few countries willing to open their markets in basic telecom services to either domestic or foreign competition. [20]  Negotiations on basic services continued,[21] but deadlocked in 1996. The impasse was rooted in the "free rider" problem inherent in MFN treatment and the ability to list exemptions:[22] countries with closed markets could maintain their systems, whilst having access to other Member's concessions and open markets, and simultaneously exclude reciprocal treatment of the latter. To this end, a set of regulatory principles crafted earlier in the negotiations were used as a blueprint to fashion market-opening

---

[18] The South African Internet Exchange. This contention forms part of an almost three-year long High Court proceeding between Telkom and the SA Internet Service Provider's Association. See *Telkom SA (Limited) v. Maepa and Others,* (8 April 1998), Transvaal Provincial Division 25840/97 (High Court of South Africa). See below note 94. Similar concerns can be raised in terms of Vodacom's holding of Internet Service Provider, Worldonline (formerly Yebonet!).

[19] See below note 96 and accompanying text.

[20] Only 48 governments were prepared to schedule basic services. See Jennifer L. Feltham, 'Polish Communications Law: Telecommunications takes off in transition countries but at what price are they becoming wired?' 33 Vand. J. Transnat'l L. 147 at 154 (2000). See also Marco C.E.J. Bronckers and Pierre Larouche, 'Telecommunications Services and the World Trade Organization' 31 J. of World Trade 3 at 6 (1997).

[21] Under the auspices of the Negotiating Group on Basic Telecommunications (NGBT). Participation was voluntary commencing with 33 WTO Members. By April, 53 Members were participating and offers had resulted from 48 governments.

[22] Article II:2.authorizes measures inconsistent with MFN if it is listed in the Schedule and meets the conditions set out in the *Annex on Article II Exemptions.* As Senunas, above note 6, points out however, these conditions are not onerous and merely call for a review of all exemptions lasting more than 5 years and provide that in principle, no exemption shall exceed 10 years.

policies,[23]and design the requisite safeguards for market access and foreign investment in domestic law.[24]

After resolving a number of negotiating issues,[25] Members finally reached consensus on the liberalization of trade in basic telecommunications services in February 1997. It is this agreement that takes the form of the Fourth Protocol[26] and augments the 1994 Schedule in exacting commitments to open markets in both basic telecommunications and satellite services.[27] Parties to the Protocol have committed over varying time frames, to dismantle the state monopoly provision of these services, open entry to foreign suppliers and adopt pro-competitive and independent regulation in the sector. Despite this commitment, monopoly legacies continue to present complexities for the introduction of competition where structural advantages for dominant players exist.[28] For this reason, the widespread adoption of the RP is arguably the most important lubricant to the mechanics of the GATS and the realization of its objectives in telecommunications trade liberalization.

Although ostensibly a guideline, the RP is in fact a substantive foundation on which to design a template for regulation. With non-discriminatory standards serving to pry markets open, the regulatory principles, woven with pro-competitive ideals, provide the means with which to keep them so. The Fourth Protocol thus recognizes that lifting formal barriers to entry is a necessary, but not sufficient condition to ensure substantive market access because of the tendency of former monopolies to dominate. Whilst the principles have been criticized as lacking in substance, I argue that they remain a solid foundation for standardizing regulatory design. This contention will be explored further below. However, to assess SA's compliance with its WTO commitments and understand the implications for domestic policy, it is necessary to briefly consider the context within which SA participated in the GATS negotiations.

---

[23] The Regulatory Paper was later adopted by all participating countries except Ecuador and Tunisia.

[24] Bronckers and Larouche, above note 20 at 23.

[25] Including the conditioning of market access on the availability of spectrum; the extent to which basic telecommunications commitments include transport of video and/or broadcast signals within their scope; the potential anti-competitive distortion of trade in international services and whether the application of accounting rates to services and service suppliers amounted to "measures" under GATS and were subject to MFN. The latter two issues particularly, remain contentious in the international telecommunications arena and will likely form an important part of negotiations in subsequent rounds. For detailed examination of these concerns, see Bronkers and Larouche, Ibid at 13. For an excellent presentation of accounting rates debate, see the full issue of 24:1 Telecommunications Policy 51 (2000).

[26] The agreement came into effect on 5 February 1998 and currently has 72 signatories.

[27] Either on 1 January 1998 or on a phased-in basis.

[28] Simply stated, this advantage arises from the fact that the former monopoly, as a dominant player, has an established network, customer base and political access to policy makers. These advantages, often result in the refusal to provide facilities and interconnection to new entrants; unfair cross-subsidization and predatory behaviour. An independent regulator is intended to restrain this tendency.

## C. Telecommunications Reform in South Africa

Telecommunications reform in South Africa needs to be located within the broader political transformation process that resulted in the demise of Apartheid and the transition to democracy.[29] In 1992, Telkom was incorporated, replacing the Department of Posts and Telecommunications (SAPT) as the vehicle through which all telecommunications services were provided.[30] The company was partially privatized in 1997 through the conclusion of a 30 per cent strategic equity partnership, valued at 1.26 billion USD.[31] This sale was driven by the need to attract the capital and management experience required to transform a debt-ridden monopoly, prepare it for competition and facilitate the goal of universal service. In exchange for the latter, Telkom was granted a five-year exclusive license on all basic voice services and the provision of facilities to all service providers.[32] Subject to a scheme of incentives and penalties, if Telkom achieved its targets by the end of 2002, its exclusivity period could be extended by an additional year.[33]

---

[29] For an excellent overview of the political and economic drivers of the reform process in South Africa, see Robert Horwitz, 'South African Telecommunication: History and Prospects' available online at http://www.vii.org (visited 8 November 2000) and Robert Horwitz, 'Telecommunications Policy in the New South Africa: Participatory Politics and Sectoral Reform' 23 Communicatio 2 (1997).

[30] The governing legislation was the Post Office Act 1958. See Myron Zlotnick, 'Telecommunications Monopoly in South Africa – some human rights aspects and options for future regulation' 43 J. African L. 214 at 215 (1999).

[31] Thintana Communications. Inc comprised of Texas-based SBC Communications Inc. (18%) and Telekom Malaysia (12%).

[32] Including public switched, national long distance and international services, local access services and public pay phone services. In the five-year period Telkom is required to rollout 2.6 million lines in total, 1.6 of these in priority areas and 120 000 new public pay phones. The so-called "Telkom License" contains three separate licenses in one Government Notice: *Government Gazette* 17984 GN R768 (PSTS); GN R769 (VANS) and GN R770 (Radio) of 7 May 1997.

[33] Telkom will not apply for the extension, which was to be negotiated between itself and the Department of Communications. This was verified by the publication of Ministerial Policy Directions on the proposed post-exclusivity market structure of telecommunications, published on 20 March 2001. Subsequent to the March draft, two policy changes occurred on the 26 July 2001 and on 15 August 2001respectively. See below note 55. While these changes affected the number of fixed line competitors to be licensed, neither proposed any variation to the date terminating the incumbent's exclusivity. See below note 65 and accompanying text. A Telecommunications Amendment Bill, published subsequent to the Ministerial Policy Directions on 29 August 2001 further confirmed the expiry of Telkom's exclusivity on 7 May 2002. See below note 57. See also Nathi Sukazi, "Calls for competition are answered, optimism grows" *Business Report,* 30 March 2001 and Nicol Degli Innocenti, "South Africa to grant two telecoms licenses" *Financial Times*, 27 July 2001, but see Shirley Kemp, "Government U-turns on telecoms decision" *Moneyweb*, 15 August 2001 online at http://allafrica.com/stories/200108160038.html.

Providing the framework for this reform, the Telecommunications Act[34] was passed in 1996, with a clearly stated public interest objective and a mandate to give effect to various important policy goals, including universal service; consumer protection; competition and innovation, growth and investment and the ownership and control of services by historically disadvantaged groups.[35] Most significantly, the Act entrenched the separation between policy formulation, operations and implementation, by establishing for the first time in SA telecoms history, an independent regulator, South African Telecommunications Regulatory Authority (SATRA).

The delivery and regulation of broadcasting services largely paralleled the history of telecommunications, but since 1993, has been regulated by the Independent Broadcasting Authority (IBA).[36] In June 2000, the Independent Communications Authority of South Africa (ICASA) Act was passed, merging the two erstwhile separate regulators into one, although retaining separate broadcasting and telecommunications divisions within the new organization.[37]

The Telecommunications Act also cemented, at least temporarily, the duopoly position of the two cellular operators – Vodacom and MTN - whose licenses were granted prior to its enactment.[38] Other non-exclusive services recognized in the legislation are Private

---

[34] Telecommunications Act, 1996. See also the Green and White Papers on Telecommunications Policy, preceding the Act, *Government Gazette* 16995 GN 291 of 13 March 1996. The latter is also available at http://docweb.pwv.gov.za/docs/policy/telewp.html.

[35] These central policy goals are expanded upon in seventeen listed objects of the Act, s 2 (a)-(q). They include *inter alia,* universal service; fair competition; economic growth and development; the ownership and control of telecommunication services by persons from historically disadvantaged groups; the promotion of small, medium and micro-enterprises; and the empowerment and advancement of women in the telecommunications industry. Similar provisions exist in the Broadcasting Act, 1999 augmenting the Independent Broadcasting Authority Act, 1993.

[36] As created by the Independent Broadcasting Authority Act*,* 1993. The Telecommunications Act does not apply to broadcasting in South Africa, except in so far as certain aspects of frequency management demand. For an overview of broadcasting in SA prior to the IBA Act, see Hudson N. Janisch and Danny M. Kotlowitz, 'African Renaissance, Market Romance: post apartheid privatization and liberalization in South African broadcasting and telecommunications' (CITI Symposium, Has Privatization Worked? The International Experience, Columbia University, 12 June 1998) (On file with the author)

[37] ICASA Act, 2000. The Council has been enlarged from 6 to 7 councilors. The Telecommunications Act and a new Broadcasting Act, 1999 continue to operate as governing legislation for both sectors. However, the Telecommunications Amendment Bill tabled in Parliament on 31 August 2001 which, when passed, will amend the principal Act. See below note 57. In this article, the regulator will be referred to as both SATRA and ICASA where necessary to indicate the pre-and post- merger timeframes.

[38] *Government Gazette* 15232 GN R1078 of 29 October 1993. The Telecommunications Act provides for the granting of additional licenses. On 25 June 2001, ICASA issued a third mobile cellular service license to Cell-C (Proprietary) Limited. See ICASA Media Release, 22 June 2001 at http://www.icasa.org.za/satra/press-page.cfm?ID=136 (visited 12 September 2001).

Telecommunications Networks (PTN);[39] Value-Added Network Services (VANS)[40] and Internet Service Provision (ISP).[41] Manufacturing and equipment supply is fully deregulated and competitive, subject only to the condition that all equipment is type approved by the Authority.[42]

Despite a vibrant and growing telecommunications sector, it needs to be noted that of the estimated nine million households in SA, only three million have telephones, with less than 20 per cent of that figure in black households.[43] In addition to joining worldwide trends in telecommunication liberalization, reform in the SA telecoms sector since 1996 has been driven by a need to redress this disparity and the racial legacies of the past that it reflects.

**III.** **SOUTH AFRICA'S COMMITMENTS UNDER THE GATS**

The GATS was signed by South Africa on 15 April 1994 and the country's offer on basic telecommunications became part of its Schedule of Specific Commitments when it entered into force on 5 February 1998.[44] The following provides a chronological outline of SA's commitments.

**A. 1994**

In signing the GATS, SA committed to open its market only in value-added or enhanced services.[45] At the horizontal level, SA made no market access undertakings and chose to remain unbound with regard to the presence of natural persons, except for the "entry and temporary stay", without an economic needs test, of certain foreign individuals in order to provide a service.[46] A national treatment limitation was entered on the local borrowing by SA registered companies with a non-resident shareholding of 25 per cent or more.

---

[39] There are two nationwide state PTNs: Transtel (a division of the transport utility, Transnet) and Esi-Tel (a division of the electricity utility, Eskom) and a number of competitive interim PTN licenses have also been granted under the Act. Two national wireless data communications licenses authorized to transmit uni or bi-directional data were also 'grandfathered' under the old regime, one of which provides the infrastructure for South Africa's first national lottery.

[40] Licensed in terms of s 40 of the Telecommunications Act, 1996. There are currently in excess of 60 interim VANS licenses; 110 Internet Service Providers (ISPs) virtual ISPs, access providers and content aggregators. A licensing framework for VANS; PTNS; Virtual Private Networks; Internet service (ISPs) and Internet access providers (IAPs) is awaiting approval by the Minister of Communications. All VANS providers, including ISPs will continue to require licenses to operate services. See *Government Gazette* 20866 of 4 February 2000 and 21642 of 11 October 2000. See above note 145.

[41] For an overview of the Internet industry in SA, see Media Africa.com, *The 4th South African Internet Services Survey,* (Johannesburg: Media Africa.com, 2000) available at http://www.mediaafrica.com.

[42] Telecommunications Act 1996, s 45.

[43] See CommUnity, a website for community Information and Communication Technology (ICT) projects in South Africa at http://www.communitysa.org.za.

[44] SA's first offer was made on 29 January 1997 (S/GBT/W/1/Add.9 (97-0335)); and then modified on 31 January 1997 (S/GBT/W/1/Add.9/Rev.1 (97-0406)) and finally re-submitted on 13 February 1997(S/GBT/W/1/Add.9/Rev.2 (97-0563)).

[45] Including e-mail; voice mail; on-line information and data base retrieval; electronic data interchange; facsimile services, including store and forward, store and retrieve; code and protocol conversion and on-line information and/or data processing (including transaction processing). WTO, (15 April 1994 ) *Schedule of Specific Commitments (South Africa)* GATS/SC/78, (94-1075).

At the specific level, SA listed a number of limitations, including the bypass of SA facilities for routing both domestic and international traffic, including callback and country direct dialing services. There was no formal policy in place with respect to VANS and Telkom remained the *de facto* regulator, authorizing all services, including service by international VANS providers, through agreements on an ad hoc and informal basis.[47]  In line with the horizontal listing, SA also remained unbound with respect to the presence of natural persons.

### B. 1997

SA's 1997 schedule of commitments on basic services[48] was fashioned from the structure and tenor of the then recently passed Telecommunications Act. In terms of the GATS horizontal listings, SA continued to impose the same restrictions with respect to market access and national treatment, as it had in 1994. The services scheduled extended to facilities based and public switched telecommunications (PSTN); mobile cellular, including mobile data and satellite services.[49] SA continued to remain unbound with respect to the presence of natural persons except as indicated above. The protection of Telkom's monopoly over voice services and facilities provision remains overarching and the legislative restriction already applying to voice over VANS was scheduled as a market access limitation.[50]

SA retained its prohibition on bypass but committed to the expiration of the monopoly and the feasibility of additional PSTN suppliers by 31 December 2003. SA also undertook to liberalize resale services between 2000 and 2003.[51] SA confirmed the mobile cellular duopoly by listing a restriction on commercial presence but committed to licensing one additional cellular operator within two years of the Protocol's commencement date.[52] Finally, commitments made in respect of fixed and mobile satellite services parallel those made for the fixed line and PSTN. SA retained a limit of up to 30 per cent on foreign investment in basic service licenses.

---

[46] For a period of up to three years, unless otherwise specified, implying a choice to permit itself flexibility in the future to restrict such entry and stay.

[47] The Post Office Act 1958, s 78(5) and s 78(6). The Schedule stated that a process was underway to introduce a regulator which might take over licensing functions and may address the lack of policy on international VANS.

[48] WTO, Group on Basic Telecommunications, *Draft Offer on Basic Telecommunications: South Africa,* GATS/SC/78/Suppl.2.

[49] Additional commitments were made in respect of paging, personal radio communication services and trunked radio system services.

[50] The Telecommunications Act 1996, s 40(2).

[51] Subject to the proviso that the Authorities would define the terms and conditions as well as the maximum limit for foreign investment.

Most importantly, while the principles that comprise the Reference Paper largely mirror the intent already expressed in the Telecommunications Act, South Africa adopted them in full. The endorsement and adoption of these principles by Members at the international level is enormously instrumental in creating consensus as to what constitutes an acceptable framework for regulation. Coupled with Article VI, which requires measures affecting trade in services to be administered in a reasonable, objective and impartial manner, this offers constructive support against any Member potentially thwarting commitments through domestic regulation.

While the commitments under the Fourth Protocol reflect a minimum undertaking, SA is free to exceed those commitments and open markets beyond that stated, as long as there is compliance with the principle of non-discrimination. Any violation of the agreement is subject to determination under the dispute settlement mechanisms of the WTO.[53]

## C. STATUS OF COMMITMENTS

### Sector Specific[54]

Since March 2001, two Ministerial Policy Directions have been published setting out the proposed market structure for telecommunications after 7 May 2002.[55] On 15 August 2001, the Ministers of Communication, Trade and Industry and Public Enterprises issued a joint statement revising the second of these policy directions and effectively setting out a third policy change in six months.[56] Shortly after that statement, a Telecommunications Amendment Bill was tabled in Parliament capturing many of the intended policy changes and proposing a number of key amendments to the existing Act. The stated aim of the Bill is to augment the legal framework for the sector, post –exclusivity and to bring the existing Telecommunications Act in line with technological, regulatory and industry developments over the past five years in South Africa and comparable international jurisdictions.[57] At the time of writing, the Bill remains open for public comment. Whilst it is still possible that it may be amended prior to promulgation, the Bill and the Ministerial Directions reflect current government policy for the sector. Where relevant, they will be discussed below. [58]

---

[52] With a commitment to undertake an economic feasibility study of a third by 31 December 1998.

[53] GATS, Part V, Article XXIII. See Thomas L. Brewer and Stephen Young, 'WTO Disputes and Developing Countries' 33 J. of World Trade 5 at 169 (1999).

[54] It is not possible to deal with the complexities of each commitment in this paper. As such, more detailed consideration of one should not imply priority of importance but rather a necessary choice in order to map out only the most current and controversial.

[55] The 20 March 2001 and 26 July 2001 Ministerial Policy Directions are available online at http://docweb.pwv.gov.za (visited 13 September 2001). Other issues canvassed include universal access and service; service rates for educational institutions; economic empowerment of historically disadvantaged persons; numbering; public emergency communications; directories and directory enquiry services.

## 1. Facilities-based and PSTN Services

In terms of its license, Telkom is authorized to provide PSTN services and facilities for a period of 25 years,[59] with an exclusive right to provision for the first five years.[60] The exclusivity expires on 7 May 2002. From that date, limited competition in basic telecommunications will be permitted with the licensing of a second network operator (SNO).[61] This will follow a public listing of Telkom on the Johannesburg Stock Exchange (JSE) in the first quarter of 2002, as part of a broader plan for the privatization of state assets.[62] The initial public offering (IPO) will see the government hand over a further 20-30 per cent of its interest in Telkom.[63]

The decision to grant only one additional fixed line service license was originally expressed in the March 2001 Ministerial Direction. It also proposed that Transtel and Esi-Tel, the telecommunication divisions of the transport and electricity utilities, were both to form part of the potential bid consortium. However, in the July 2001 Direction, citing consultations with the industry as the main reason for revision, the Minister of Communications announced that a third network operator (TNO) would also be licensed and that Transtel and Esi-Tel were to be included separately in both respective bids.[64] Yet in the August 2001 joint policy statement, the Minister announced a return to the original policy of only one additional network operator, with a TNO being licensed in 2005, subject to the completion of a market feasibility study.[65] Again, provision is made to include a state player in the domestic investment component: either Transtel or Esi-Tel but not both, as originally suggested in the March 2001 policy.[66] The SNO will receive a full service license and be able to compete with Telkom in international, national long distance, payphone, local access and value-added network services. Consistent to all three versions of the proposed policy, up to 30 per cent of equity will be set aside for black economic empowerment. Finally, the March and July 2001 proposals indicated that foreign shareholding in major licenses will be restricted to 49 per cent. However, in the August statement, all foreign ownership restrictions were lifted, although the Ministers did not exclude the possibility that such a restriction could still be specified in the invitation to apply for the license.[67]

---

[56] This policy revision has not been formally gazetted.

[57] [B65-2001]. The Bill was published as *Government Gazette* 22630 of 29 August 2001 and is available online at http://www.parliament.gov.za/bills/2001/index.htm (visited 14 September 2001). The Bill also includes a number of measures necessary to facilitate the Telkom IPO, discussed below. See note 63 and accompanying text.

[58] Many of the proposed amendments have no direct bearing on the subject matter of this paper and for this reason, detailed discussion of the Bill will not be included here.

[59] Telecommunications Act 1996, ss 36(1)(a), 36(7)(c) and 36(9)(a).

[60] Section 36(1)(a) and 30(3)(a).

[61] See above note 33.

[62] The others are Eskom, Transnet and the Airports Company, all in varying stages of privatization. Telkom's assets exceed R27 billion, making it the country's third largest parastatal after Eskom and Transnet.

[63] The IPO is valued at between R80-100 billion (9.2–11.6 USD) and is expected to be one of the largest ever on the JSE. 'Telkom IPO among the biggest' *News24.co.za,* 23 October 2000 at http://www.News24.coza. A further 3% stake in Telkom, estimated to be worth more than R600 million was recently awarded to Ucingo Investments, a black economic empowerment group. The deal is effective from 23 March 2001. 'Ucingo wraps up stake in Telkom' *Business Report,* 3 April 2001.

---

[64] Robyn Chalmers and Lesley Stones, "Minister loses her voice and escapes some tricky questions", *Business Day*, 27 July 2001.

[65] Speculation fuelling the policy reversal suggests that intensive lobbying by Telkom and M-Cell, the parent company of cellular operator MTN and a potential bidder for the SNO is responsible for the turnaround. Government holds a 24.5 per cent stake in M-Cell, who lost R6 billion on its market capitalization following the second policy announcement in July 2001. Looking to sell its stake in the company, licensing only one competitor to Telkom is presumed to facilitate government realizing a higher value for its shares. See Lesley Stones, "Industry plea for state to rethink telecoms policy", *Business Day*, 31 July 2001 available at http://allafrica.com/stories/200107310141.html. Other reasons cited include the potentially adverse impact and reduced investor interest that two fixed line competitors might have on the proposed Telkom IPO. The July 2001 proposals to license two operators raised concern that Thintana Communications Inc. might sell a third of its 30 per cent stake in Telkom following the IPO, in protest against the government's telecommunications policy. Due to its shareholders agreement, this would limit the amount of equity that government could sell in Telkom, which might severely damage the listing and affect government's efforts to reach its revenue targets from privatization. See Robyn Chalmers, "Telkom's listing may be in jeopardy" *Business Day*, 31 July 2001 and also, "Lack of clarity may harm Telkom listing" *Business Day*, 30 July 2001.

[66] See Solomon Mokgale 'SNO storm is brewing over stakes in new operator' *Business Report,* 11 April 2001 and Nathi Sukazi, 'Transtel, Esi-Tel pine for 50% of second network' *Business Report,* 16 March 2001. See also "We Will Not Fight Eskom for SNO, Says Transtel" *Africa News,* 17 August 2001.

[67] Philip de Wet, "Third national operator, broadband abandoned" *ITWeb*, 16 August 2001. In a recent joint Communications Portfolio and Labor Select Committee briefing, the Minister noted that flexibility was necessary given the industry's state of flux. Government had therefore not wanted to specify how much of the market foreign companies could own as changing this figure at a later stage would be difficult if it was entrenched in law. She also noted that the specifics could easily be dealt with in shareholders agreements, voting pools or licensing arrangements. Also at this briefing, the Minister of Communications defended accusations that the policy process has been a 'flip-flop', stating that versions previous to August 2001 were merely drafts and not the Policy Direction

Sentech (Pty) Ltd, the state-owned signal distributor, will have its current license amended to enable it to compete only in international services. Originally the policy envisaged that Sentech could offer service direct to the public, but in what has been billed as a 'clarification' rather than a revision, the August statement anticipates that the parastatal will be issued with an international gateway license, but it will be a "carrier of carriers" and will not provide service direct to the public.

In spite of a number of administrative concerns regarding the legislative fiat with which these proposed license amendments will occur, the policy proposals in this area accord with SA's commitments under the Fourth Protocol. Notwithstanding these questions, it is doubtful however, whether any meaningful competition or material impact on price for consumers will eventuate, given this clear attempt to secure telecoms revenue for the government through the award of licenses with continued, and in some cases, substantial state shareholding.

In respect of the commitment to introduce resale between 2000 and 2003,[68] progress is arguably stunted. The White Paper contemplated resale in 2001, as part of an ideal to encourage new and innovative uses of technology and gradually open market segments to competition.[69] No corresponding provisions however materialized in the 1996 Act. The Policy Directions remain at best, equivocal on the matter. The March 2001 version contemplated services-based competition during the last quarter of 2005. The July 2001 version cited its introduction in 2004. Although somewhat unclear in its operation, the Telecommunications Amendment Bill appears to indicate the licensing of a services-based competitor to Telkom and the SNO only in 2005.[70]

Discussion on resale was absent during the early part of the exclusivity, presumably to preserve revenue flows from voice services to Telkom and more recently, in order to maximize Telkom's IPO. As a result however, it is arguable that SA is potentially breaching its Fourth Protocol undertaking to introduce resale by 2003.[71] Perhaps more significant for consumers than

---

itself. The minutes of this meeting are available online at http://www.pmg.org.za/viewminute.asp?id=844 (visited 15 September 2001).
[68] Subject to the proviso that the Authority would define the terms and conditions as well as the maximum limit for foreign investment.
[69] Clause 2.6. Chapter II of the White Paper, above note 34.
[70] Section 32A(1).
[71] Presumably, the government may argue that they have complied with resale obligations by introducing a SNO and other competitive licenses. Further, that the commitment does not specify whether resale services are voice or data, and that technically data resale is permitted, implicit in the provisioning of VANS services. This contention would

a potential GATS violation, the existing prohibition on VANS offering voice services and voice over IP (VoIP) remains in force.[72] This unfortunate policy decision is likely to have a negative impact on the universal service imperative and remains a lost opportunity to introduce competition and lower costs.[73]

## 2. Mobile Cellular Services

Under the Fourth Protocol, SA committed to licensing one further cellular operator within two years, and the completion of a feasibility study by 31 December 1998. This undertaking too, echoed provisions already contained in the Act.[74] The preferred bidder was confirmed on 16 February 2001,[75] over a year later than the date undertaken in SA's commitments.[76] This award

---

be difficult to make, given that Telkom claims such resale is illegal under its exclusivity. I argue that in the absence of a definition to limit resale to data, the term must be interpreted in its common industry usage, that is, the sale to end consumers, of bulk discounted voice minutes purchased by resellers from network operators. Finally, it is apposite to note that this commitment was scheduled as an additional one, making it a positive undertaking.

[72] Telecommunications Act 1996, s 40(2).

[73] Given Telkom's most recent tariff hike in excess of 16% for local calls, one has to seriously question the wisdom of this choice. Despite the argument that this increase may be merely part of the rebalancing which must precede competition, it has been characterized by ICASA as having an adverse effect on consumers and "potentially having a negative impact on access to telephony." See 'Reluctant Green Light for Telkom Hikes' *Mail and Guardian,* 8 December 2000. It should however be noted that the July 2001 Policy Direction provides for a Ministerial Policy Review after two years, to determine the feasibility of extending VoIP to other operators in the telecoms industry. See clause 6.4.

[74] Section 27(9) and s 34. For an account of the background, procedure, evaluation criteria and reasons for the announcement of Cell C as the preferred bidder, see SATRA, Media Briefing, "Council Intended Recommendation in Terms of section 35(1)(a) of the Telecommunications Act No. 103 of 1996 in respect of Applications for the Third Mobile Cellular Telecommunication Service License", 29 February 2000, at http://icasa.org.za

[75] 'Decision of the Minister of Communications on the Third Mobile Cellular Telecommunications License', Press Release, issued by the Ministry for Communications, 16 February 2001, at http://docweb.pwv.gov.za.

[76] The feasibility study, an invitation to apply and consequent application hearings, with bids received from international and local consortia were held within the specified time frame. Bids were received from 9 companies including AfricaSpeaks, Nextcom Cellular and Telia Telenor. Substantive delay in the process began with the announcement of Cell-C as the preferred winner in July 2000, triggering a spate of court proceedings on the basis of the concerns set out above. Losing bidder NextCom fired the first salvo in applying for an urgent interdict to prevent the Minister from confirming Cell C as winner. On the basis of an affidavit by the outgoing SATRA Chairman that alleged government interference in Council deliberations, the court granted the injunction, pending judicial review of the bidding process. Other grounds for the order include SATRA's alleged failure to consider independent assessments of the bids. In September 2000, the Minister sought leave to appeal in the Pretoria High Court on the basis that she was drawn into the licensing process to ensure that SATRA adjudicated in a fair and transparent manner. This, she claims was misinterpreted by rival bidders and the court as executive interference. The Minister also threatened to take the matter to the Constitutional Court for a decision as to whether the High Court in fact had the power to prevent her awarding the license before she had officially named the winner. Nextcom later withdrew only that part of the interdict preventing the Minister from announcing the winner, amidst speculation that a deal between Cell C and Nextcom had been struck. See Leslie Stones, 'Cellular License Brawl Continues' *Business Day,*11 September 2000; Luke Baker, 'NextCom to sue if Cell C wins cell license' *Mail and Guardian,* 5 April 2000 and Hilary Gush, 'South African Mobile License Saga Takes New Twist' *Reuters*, 18 September 2000.

followed months of controversy, exposed serious internal discordance within the regulator,[77] raised allegations of improper tendering procedures,[78]corruption, government interference,[79] and undisclosed links to applicant bids.

Coupled with a spate of resignations, court proceedings and a merger of the two regulators during the protracted licensing process, the third cell debacle has arguably been the single biggest blow to ICASA's credibility and perceptions of competency.[80] It has also raised international concerns with regard to further foreign investment in South Africa. Whilst many factors would arguably mitigate against a WTO finding of non-compliance, issues raised regarding regulatory independence have endured following a court review of the SATRA tender selection process.[81] This can be linked in part to a curious aspect of SA's regulatory framework: the power of the Minister, as opposed to the regulator, to invite applications for services and to grant licenses.[82] This has aptly served to reflect the difficulties created by having this role split between an executive arm of government and an independent agency.[83]

These licensing provisions and the implications for independence were questioned by international advisors hired to assist with the merger of the two authorities. It was suggested in this regard that SA amend its Act to effect compliance with its WTO undertakings.[84] As new and arguably more intricate roles for the Minister and the regulator in license decisions and awards appear to be contemplated in the Telecommunications Amendment Bill, remedial action should be taken to avoid a similar risk occurring in the licensing of the SNO.[85]

---

[77] Ivor Powell, 'Satra split over Cell C' *Mail and Guardian,* 31 March 2000.

[78] It was alleged that SATRA ignored a BDO Spencer Steward analysis report that did not recommend Cell C due to an inadequate business plan. SATRA mandated a consulting firm, Grant Thornton Kessel Feinstein (GKTF) to audit the decision, however controversy erupted when GTKF later disclosed previous contact with a bidder. The Communications Minister threatened to sue GTKF and the losing bidder, Telia-Telenor threatened to sue SATRA and government for the way in which the process had been handled. See Ellis Mnyandu 'Saudi-Backed Phone Group Defends South Africa Bid Win' *Reuters,* 28 June 2000, available at http://www.kagan.com/archive/reuters/2000/06/28/2000062817mma.html. (visited 9 December 2000)

[79] During deliberations, SATRA Chairman Nape Maepa recused himself under allegations of pressure from the President's Office.

[80] It has also been financially disastrous for applicants, with Cell C alone having spent more than R2 million in legal fees. Other costs are to investor confidence, job creation, black economic empowerment, real price and service competition to existing operators and more than R4 billion in peak investment, of which half would have been foreign direct investment. Marina Bidoli, 'Debacle puts the frighteners into telecoms investors' *Financial Mail,* 6 October 2000.

[81] Phillip de Wet, 'Cell C to rollout, Nextcom fights on' *ITWeb,* 19 February 2001at http://www.itweb.co.za/sections/telecoms/2001/0102190724.asp?O=Eue. ICASA has opposed NextCom's application that ICASA's predecessor, SATRA, acted in bad faith in relation to the licensing process and that ICASA is to pay the costs of the High Court proceedings.

[82] s 34(2)(a)-(b) read with s 35(2).

[83] s 34(2)(a)(ii).

[84] Alan Darling, former Canadian Radio, Television and Telecommunications Commission (CRTC) Executive Director, questioned this issue. See Marina Bidoli, 'We're Breaking Up' *Financial Mail,* 23 April 1999.

*3. Satellite-based Services*

Despite competition in Global Mobile Personal Communications (GMPCS) originally having been viewed as a priority by the Department of Communication and SATRA, a finalized policy is still not in place.[86] Satellite communications for switched telecommunications is currently provided on a monopoly basis through Telkom. Exclusivity is assured through its license, which provides Telkom exclusive rights over all international service.[87] There is no additional commitment on satellite-based services, but SA has undertaken to schedule an undertaking within one year of adopting measures in this area.[88] New, albeit draft policy in this regard reinforces Telkom's right to operate and manage GMPCS gateway stations and to provide GMPCS gateway services in South Africa.[89] In terms of the commitments under the Fourth Protocol, competitive provision in satellite services must be established by 31 December 2003.

**REFERENCE PAPER ON REGULATORY PRINCIPLES**

Many commentators note that the obligations undertaken through the Fourth Protocol and the RP are mere reflections of Members' existing or planned domestic liberalization policies. The RP's real strength however, lies in the fact that it ensures that this implementation occurs according to a set time frame. If one considers the delays and conflicts already besetting SA's telecoms sector, the importance of this document, as an ongoing facilitation tool cannot be overstated.

---

[85] See sections 34, 35 and 35A.

[86] In December 1998, the Minister issued a Policy Direction on GMPCS in South Africa. See *Government Gazette* GN R 1609 of 4 December 1998. In early 2001, a notice of intention to amend the draft policy necessitated by recent jurisprudence, was promulgated. See *Government Gazette* GN R4609 of 8 December 2000. Due to space constraints, the implications of the proposed amendments cannot be discussed here but do not have an impact on the above discussion given that the policy is still not yet finalized and is subject to possible amendment in light of the Ministerial Policy Directions of 2001 and the Telecommunications Amendment Bill.

[87] Telkom License, provision 3.1(b).

[88] Laura B. Sherman, "Wildly Enthusiastic' About the First Multilateral Agreement on Trade in Telecommunications Services' 51 Fed. Comm L.J. 61 (1998).

[89] Clause 14.1.1. See Luisa Dos Santos, 'Telecommunications: Satellite Services' Industry Series Analysis, U.S. & Foreign Commercial Service and US Department of State, 1999. Also available on Industry Canada's website at http://info.ic.gc.ca/cmb/welcomeic.nsf/icPages/Menu-e.

Structurally, the RP is divided into six sections. The first two apply to regulation of "major suppliers",[90] while the remaining four deal with general regulatory issues.

*1. Competitive Safeguards*

The RP requires "appropriate measures" to prevent major suppliers from engaging in or perpetuating anti-competitive practices,[91] such as cross subsidization;[92] improper use of information obtained from competitors,[93] and the withholding of technical and commercially relevant information about essential services. Although not the subject of WTO action,[94] it is noteworthy that that the dispute between the Internet Service Providers Association (ISPA) and Telkom, which started in 1996, contained allegations of two of these three practices.[95] Bronckers and Larouche suggest that an appropriate accounting system, with regular reporting and disclosure requirements, is the only effective mechanism with which to monitor these practices.[96] A continued problem in this regard for SA ISPs in terms of their claim is the fact that the obligation on Telkom to prepare regulatory accounting does not take effect until the end of year five of the monopoly.[97] In the interim, Telkom is authorized to cross-subsidize within its basic basket of services.[98] Without an accounting requirement in place, it remains impossible to determine whether these revenue flows seep out into other areas of non-exclusive provision, such as Internet services and VANS.

---

[90] A "major supplier" is one defined as one that can materially affect the terms of participation (price and supply) in the relevant market for basic telecoms services as a result of its position in the market, or its control over "essential facilities," (facilities of a public telecoms transport network or service that are exclusively or predominantly provided by a single or limited number of suppliers, and cannot feasibly be economically or technically substituted in order to provide a service).

[91] For an excellent assessment of competition in telecommunications and international trade in light of convergence, see Arlan Gates, above note 12 at 83. See also Marcus Fredebeul-Krein and Andreas Freytag, 'The case for a more binding WTO Agreement on regulatory principles in telecommunications markets' 23 Telecommunications Policy at 628 (1999).

[92] Not all cross-subsidization is anti-competitive, but it can easily become so in monopoly dominated markets when the operations from the profit making area are used to undercut competitors prices in non-exclusive service sectors. See Bronckers and Larouche, above note 20 at 27.

[93] For example, information obtained through interconnection negotiations or where the major supplier requires full technical and other information to be lodged for a service order.

[94] The Fourth Protocol had not yet been signed and the market had not reached the same level of maturity with the presence of foreign providers as it has now.

[95] In the initial complaint to the then Competitions Board who had jurisdiction prior to the formation of SATRA, the ISPA alleged that Telkom was providing SAIX with certain advantages that were anti-competitive. For example, the unfair cross-subsidization between voice and Internet services, fuelled by the fact that SAIX could offer dial-up services at a substantially lower rate than the private ISPs; free co-location for SAIX in Telkom exchanges and rapid responses to line applications for SAIX on behalf of clients, for which ISPA members were often subject to long delays. Finally, that SAIX was engaging in predatory pricing and piracy of ISPA's customer base as information about the prospective client, including contact details, had to be submitted to Telkom on application for a leased line. For a detailed account of the ISPA claims to the Competition Board, see http://www.ispa.org.za/submission.html, and for a more detailed history of the relationship between Telkom and the independent ISPs, see D. Kotlowitz, 'Telkom, South Africa's Internet Anschluss: A Cautionary Tale" (1998) [Unpublished, archived at the University of Toronto] (On file with the author).

[96] Bronckers and Larouche, above note 20.

Nonetheless, many commentators have criticized the RP provision as "toothless", most notably for the failure to specify a mechanism with which to address such practices, leaving members to determine their own, non-standardized measures.[99] This, it has been argued, has two adverse implications: the first, arguably less plausible one is that the vague definition of a 'major supplier' can be read to include foreign operators who are major suppliers in their own home country and measures can thus be directed at them, impeding market entry. The second suggests that the lack of safeguard rules for governments when adopting the regulatory framework, might lead to a failure in implementing effective action for preventing anti competitive measures by domestic suppliers with market power.[100]

There are however important policy reasons why Telkom is not subject to regulatory accounting at this point and to the extent that this is in fulfillment of an important domestic policy goal - the attainment of universal service - it is arguable that the WTO has no place in imposing strictures. While there have been some questions raised about Telkom's separation of revenue between competitive and exclusive service sectors, there is no glaring evidence suggesting that it has indeed violated these provisions.

---

[97] Clause 8.1 of the Telkom PSTS license. This is to be established in accordance with the Chart of Accounts and the Cost Allocation Manual (COA/CAM).

[98] In terms of clause 9.7 read with 3.1 of the Telkom PSTS License, tariffs need only be rebalanced at the end of the exclusivity period. For consideration of tariff rebalancing in the SADC region, see Southern Africa Telecommunications Restructuring Program, "Study Report: Telecommunications Tariff Rebalancing", A SATCC Telecommunications Sector Development Program, funded by USAID, 12 November 1998. As the report aptly points out, tariff rebalancing towards cost is crucial in gradually liberalizing markets as experience shows that international and long distance calls will be the first area targeted by new suppliers for their profit margin. Where tariffs are cost-based, competition will be distorted and the overall objectives for telecommunications development may be undermined or defeated, at 18.

[99] Fredebeul-Krein and Freytag, above note 91, and Sherman, above note 88. But for a defense of this principle, see Chantal Blouin, 'The WTO Agreement on Basic Telecommunications: a reevaluation' 24 Telecommunications Policy 135 at 139–40 (2000).

[100] Fredebeul-Krein and Freytag, Ibid, at 628-9.

*2. Interconnection*

Interconnection principles arguably constitute the core of the RP as competition is totally precluded in the absence of an effective policy in this regard.[101] Safeguards ensuring interconnection on reasonable terms and within a reasonable time are crucial to allow new market entrants access to the network of an established or monopoly provider.[102] Efficient interconnection requires certain common elements, enumerated in the RP, reflecting general principles adopted domestically by many liberalized countries.[103] There are also some shortcomings to the principles as articulated in the RP, such as the failure to specify a costing basis, the degree to which networks have to be unbundled and whether a regulator can enforce an interconnection agreement. The RP does, however, suggest some baseline guidance.[104] Most significantly, a transparency requirement that both the procedures for interconnection and any concluded agreements of major suppliers must be publicly available applies. With this in mind, the status of interconnection policy in South Africa has raised concerns and also serves to reflect the tensions existing within the regulatory environment.

The Telecommunications Act requires mandatory interconnection of Telkom's network on request, unless such interconnection is unreasonable.[105] Ministerial Guidelines governing Telkom in this regard, lapsed in May 2000.[106] The Act requires the Authority to prescribe guidelines for the industry relating to the "form and content" of interconnection agreements and to determine fees, service levels and time frames.[107] In March 2000, with Ministerial approval, SATRA published interconnection and facilities leasing guidelines,[108] which were well received by an industry bereft of recourse against Telkom from the commencement of the Act, for continued failure or delay in providing facilities to non-exclusive service suppliers.[109] A month later, the Minister published a Government Notice unilaterally withdrawing them, on the basis that there had been insufficient public consultation and that the merger between the IBA and SATRA required further postponement for the IBA to participate in the process.[110] SATRA reinstated the guidelines, claiming that the Minister's actions were *ultra vires* and resolved that it would continue to apply the withdrawn guidelines.[111] Aside from public criticism of SATRA's actions, the Ministry remained silent on the issue.[112]

---

[101] Clause 2.1. For an excellent account of interconnection issues, including Internet interconnection and the implications for interconnection on accounting rates, see ITU, *Trends in Telecommunication Reform, 2000-2001: Interconnection Regulation.* An overview of this document is available online at http://www7.itu.int/treg/Events/Seminars/2000/Symposium/English/Document14_tim.pdf (visited 12 December 2000). See also W.H. Melody, 'Interconnection: the Cornerstone of Competition' *Telecom Reform: Principles, Policies and Regulatory Practices* (Lyngby, Denmark: Den Private Ingeniorfond, Technical University of Denmark, 1997).

[102] Fredebeul-Krein and Freytag, above note 91 at 629.

[103] Ibid. Interconnection should be provided on request, be equitable and non-discriminatory in terms of quality; provided at technically feasible points in the network; timely and on reasonable terms and conditions, cost-oriented and sufficiently unbundled so that new entrants are not paying for unnecessary network components and facilities.

[104] Ibid, at 630. I disagree that the latter claim is indeed a weakness. It seems quite clear that 'recourse to' must surely imply an enforcement measure. A domestic interconnection policy that makes no provision for agreements to be imposed following procedures and arbitration would arguably be falling short of both international standards and the goal behind the inclusion of the principle in the RP.

[105] Section 43(1)(b) and (c). 'Reasonableness' is determined with reference to 'technical feasibility' and whether "interconnection will promote increased public use of the telecommunication system, or more efficient use of telecommunication facilities."

[106] *Government Gazette* 17984 GN R771 of 1997, Ministerial Determination on Interconnection Guidelines.

Although the High Court has ruled these Guidelines to be of full force and effect,[113] it was within this context that a number of disputes between various service providers and Telkom, were to be heard by the regulator: if the guidelines were in fact void, the regulator was once again in limbo; if they were applied by SATRA, in the face of their withdrawal by the Minister, Telkom refused to acknowledge any legal validity to them or the processes under which they were being invoked.[114] It is also within this context that the USTR threatened to take SA to the WTO.

---

[107] Section 43(3).

[108] *Government Gazette* 20993, GN R1259 of 2000, Interconnection Guidelines issued by the Authority in terms of Section 43 of the Telecommunications Act 1996', 15 March 2000. The Draft Guidelines are available on SATRA's website at http://satra.gov.za/publications.htm. This process was preceded by public consultation and hearings held on 13 and 14 November 1998. See *Government Gazette* 19159 GN R1683 of 1998.

[109] It is of course arguable that the provisions of section 43 and 44 of the *Telecommunications Act* contain sufficient authority for the regulator to act in the case of failure to supply facilities and refusal to interconnect where the objective criteria are satisfied, and that the Guidelines do not constitute the only authority to do so. Moreover, the public interest objective threaded through the Act coupled with a general enabling provision in section 5(1)(b) that authorize the regulator to perform any acts necessary for the performance of its functions, should provide additional grounds on which to intervene. However, SATRA repeatedly refused to act on a range of facilities leasing disputes in the industry until such time as the guidelines were in place.

[110] *Government Gazette* 21107 GN R1680 of 2000.

[111] SATRA, Media Release, 6 June 2000 at http://satra.gov.za/press-page.cfm (visited 12 December 2000). See also 'Satra's Last Stand' *World Reporter,* 6 June 2000.

[112] The Department of Communications criticized the regulator's behavior as "the last kick of a dying horse" and the actions of those who "did not have the interest of government and SA's telecoms industry at heart," See, 'Ngcaba slammed for Comments', *ITWeb,* 9 June 2000 and Lesley Stones, 'Clash Looms as SATRA takes on Minister' *Business Day,* 8 June 2000. See also Lesley Stones, 'Defiant SATRA says only a Court can reverse its move', *Business Day,* 12 June 2000.

[113] *Telkom SA Ltd* v. *The Independent Communications Authority of SA et al*, Unreported, Case No.: 19014/2000, 19 March 2001.

**The WTO Dispute: US and SA[115]**

By virtue of its predecessor's operations, AT&T is a deemed VANS license holder under the Telecommunications Act.[116] In terms of the Act, all license holders are required to use Telkom infrastructure for the provision of their services, until the termination of the latter's monopoly.

In July 1999, Telkom began to request written confirmation from VANS providers that they were not using their facilities for the provision of any service in violation of Telkom's exclusivity.[117] When Telkom denied further requests for facilities, the SA Association of VANS providers (SAVA), of which AT&T is a member, approached SATRA for assistance. In September, the SATRA Council declared Telkom's behavior to be anti-competitive, but stopped short of ordering them to meet outstanding service requests. The fact that the interconnection and facilities leasing guidelines had not yet been promulgated was a further impediment as SATRA claimed it was prevented from granting relief under the Act.[118] SAVA petitioned the High Court which referred the matter back to SATRA for determination, claiming that it lacked the competence to decide on the technical nature of the dispute, and further that it could make no determination while SATRA was seized of the matter.[119] This ruling caused both parties to question SATRA's competence once again: SAVA claimed that SATRA's inability to adjudicate was the reason why they sought High Court intervention in the first place, and Telkom claimed that SATRA's continued lack of response to its complaints regarding VANS providers since 1997, had forced Telkom to begin withholding facilities.[120] The fact that SATRA had made little progress on finalizing a VANS licensing framework was also a contributory factor as the industry lacked certainty regarding the status and ambit of services included under a VANS license.[121]

---

[114] In this context, an interim ruling on an interconnection dispute between Telkom and Wireless Business Solutions (Pty) Limited (WBS) was handed down by SATRA on 11 February 2000. This ruling, recently challenged in the High Court, imposed terms and conditions for an interconnection agreement between Telkom and WBS. Both parties provide telecommunication facilities to the newly licensed, national lottery operator, Uthingo. Telkom claimed that the Interconnection Guidelines used by SATRA to make the ruling were invalid as the Minister had withdrawn them; that improperly followed delegation procedures by SATRA rendered their decision null and void; and that the terms of the agreements imposed on them through this process were invalid. The court upheld Telkom's latter two claims, but dismissed the first one. Other matters formed part of this dispute, including a complaint by WBS in respect of alleged anti-competitive actions by Telkom with regard to pricing in respect of Swiftnet, a Telkom's subsidiary competing with WBS. Despite the importance of this case as the first interconnection dispute to be contested in a SA court, it is highly complex and more appropriately the subject of another paper entirely.
[115] While the potential dispute between the USA and SA over Telkom's refusal to supply facilities, is largely a "competitive safeguards" dispute, the lack of clarity over the facilities leasing guidelines makes it suitable for consideration in this context.
[116] Section 40 deems any person who was providing a VAN service in terms of agreements with Telkom under section 78(2)(a) of the Post Office Act, immediately prior to the commencement of the Telecommunications Act, to be the holder of a license to provide that service. AT&T's predecessor, Trafex (Pty) Ltd was trading under those conditions since 1985.
[117] Jennigay Coetzer, 'VAN Service Providers Complain of pressure' *World Reporter,* 6 October 1999.
[118] The relief sought was under s 44(7) of the Act which provides in relevant part that, "where the Authority is satisfied that Telkom is unwilling or unable to make suitable facilities available to that person within a reasonable period of time, the Authority may, instead of proposing terms and conditions… authorize that person to provide or obtain any necessary telecommunication facilities other than from Telkom on conditions determined by the Authority." See above discussion note 109.

At this juncture AT&T, frustrated with lack of recourse at the domestic level, approached the USTR for assistance. Including SA in its annual section 1377 review of telecom trade barriers in other countries, the USTR set a June 2000 deadline for "co-operative resolution of the dispute" failing which they threatened to take action under the WTO's Dispute Settlement provisions.[122] The claim suggests that Telkom's refusal to provide facilities is contrary to SA's WTO obligations to provide market access and national treatment for VANS.[123] It is further claimed that SA is obligated to prevent Telkom from abusing its monopoly position when competing in the supply of a service outside the scope of its monopoly rights. [124] Finally, that under the provisions of the Annex, SA is required to ensure that US VANS suppliers receive "access to and use of public telecommunications transport networks and services on reasonable and non-discriminatory terms and conditions."[125]

---

[119] *AT&T Global Network Services SA (Pty) Ltd et al,* v. *Telkom SA Ltd,* Unreported Case No.: 27624/99, 28 October 1999.

[120] Lesley Stones, 'Court sends Telkom dispute to SATRA' *Business Day,* 4 November 1999. Telkom claims that facilities sharing between VANS and their clients is costing Telkom between R500 million -R1 billion annually, because companies are offering services to their clients that should have come from Telkom.

[121] See below note 145.

[122] Under s 1377 of the Omnibus Trade and Competitiveness Act, 1988. There were nine countries under review in 2000 in addition to SA; including Israel (discriminatory access fees on calls to and from Canada and the US); Japan (failure to implement cost based interconnection charges); Germany (backlog of interconnection requests unfilled and high license fees); the UK (unbundling and line sharing for DSL services); Taiwan (preservation of exclusivity and high license fees); Canada (concern over the contribution regime for international providers) and Peru (non cost-oriented and discriminatory interconnection); Mexico (universal service rules and interconnection and international service regulations. See Office of the United States Trade Representative, Washington, D.C., *Annual Review of Telecommunications Trade Agreements Highlights Concerns Regarding Mexico, South Africa, and Other Countries*, Apr. 4, 2000. The 1377 Reports for 2000/1 are available at http://www.ustr.gov/sectors/industry/telecom.shtml.

[123] Article XVI and XVII respectively and Article VIII:1.

[124] Article VIII:2

[125] Provision 5(a), above note 8.

After re-hearing the matter under the court's order, SATRA directed Telkom to resume the supply of facilities to VANS operators and to complete all backlog orders within forty-five days.[126] SATRA also reacted with some degree of opprobrium to Telkom assuming regulatory functions and determining for itself the legality of use for which the facilities were being requested.[127]

In June however, the USTR withdrew their threat of WTO action as out-of-court negotiations between the parties resulted in Telkom agreeing to restore AT&T's access to its network.[128] Telkom have since filed a separate complaint with ICASA alleging that AT&T is providing services outside the scope of its VANS license and has once again ceased the supply of facilities to AT&T.[129] This issue has again been cited in the US 2001 annual 1377 review for appropriate action. Threats of such action, whilst also subject to institutional delays serves to signify a serious and rising sense of dissatisfaction with domestic regulatory and enforcement structures. Part III will consider the effectiveness of seeking redress at the WTO in light of the many criticisms leveled at the Fourth Protocol.

*3. Universal Service*

Although GATS Members have committed to a common vision on liberalization in signing the Protocol, it is evident that within the divergent country Membership, non-economic policy objectives, often contrary to the GATS ideals, will be pursued. These cannot always be quantified in empirically measurable ways, and will mean different things to different countries at different times. The adoption of a universal service policy is usually one such goal. Depending on where the policy is instituted, (developing or developed country) and the level and origin of contribution to it (firms or subscribers) and the model chosen by government to effect it (monopoly or multiple providers) will inform the degree to which it is valued. Precisely because access to telecoms infrastructure among Members is so uneven, the 'value' of such a policy can never be equally quantified across the spectrum of GATS signatories. To this end, GATS recognizes Members rights to regulate services in order to meet divergent national policy objectives. GATS however requires that any measures pursuant to these objectives be carried out in a transparent, non-discriminatory and competitively neutral manner.[130] For example, Canada's subsidization of local phone service to remote and high-cost areas, was cited by the USTR as anti-competitive for certain long-distance carriers. [131] No recourse at the WTO was sought however, in light of the CRTC's recent (and arguably unrelated) efforts to reform the contribution program.[132]

---

[126] SATRA Judgement in the matter heard on the 6 and 7 June 2000 between the SAVA and Telkom SA Ltd. Available on SATRA's website at http://satra.gov.za.

[127] Ibid, Para 30.

[128] 'USTR Decides to keep monitoring Germany, South Africa, UK' *Washington Telecom Newswire,* 16 June 2000. New allegations of anti-competitive behaviour have begun to emerge. Recently, some SAVA members went on record that Telkom has threatened not to provide companies with additional bandwidth unless they sign a three-year contract, which will go beyond Telkom's exclusivity period. See Marina Bidoli, 'Flurry of Activity as market gets ready for selloff" *Financial Mail,* 15 September 2000.

[129] 2001 Section 1377 Review of Telecommunications Trade Agreements, 2 April 2001.

In SA, an overarching objective of the Act is the supply of telecommunications services to those who were deliberately excluded by Apartheid.[133] The White Paper characterized affordable communications for all as "the core of its vision and is the goal of its policy."[134] This is articulated in the objectives of the Act, and in the establishment of a Universal Service Fund (USF) which will receive contributions from all licensees.[135] The basis and manner of the contribution and the dates for payment are to be determined by ICASA and prescribed by regulation.[136] Although this has not yet been finalized, the mechanisms anticipated are thus far GATS compliant: any measures adopted will apply to all service suppliers under all licenses in a similar manner and under a prescribed formula.

---

[130] Reference Paper, Para 3, See also Fredebeul-Krein and Freytag, above note 91 at 631.

[131] Above note 122.

[132] See Decision CRTC 2000-745, Ottawa, 30 November 2000, Changes to the Contribution Regime, Reference: 8695-C12-06/99 at http://www.crtc.gc.ca/archive/Decisions/2000/DT2000-745e.htm. One cannot be sure that the CRTC was responding to WTO concerns in this process. For an overview of the implications of the GATS Fourth Protocol on Communication Policy in Canada, see Janisch, above note 17 at 65-86.

[133] While the objectives of the Act are not placed in a hierarchy, it is arguable that redressing apartheid exclusions is the de facto overarching one. Telkom's exclusivity is predicated on the assumption that the state is best placed to deliver services to unserved areas, that would not otherwise be serviced by competitors in a free market, who would "cherry-pick" only the profitable aspects of that market. Most conflict in the industry stems from Telkom's obligations in this regard: Telkom claims it is merely defending its roll-out plan, and government's important national policy, and competitors suggest that Telkom uses this claim as disguised protectionism. See above note 35.

[134] Clause 1.1. The current national teledensity figure per household is approximated at 42 per cent overall, defined as the percentage of people who have a phone, cellular or fixed, in the dwelling where they live. The overall national universal access indicator, defined as the percentage of people who have access to a telephone, stands at 80. See above note 43.

[135] Sections 65-67. The USF will be used exclusively for the payment of subsidies to assist needy persons towards the costs of telecommunications services. See *Government Gazette* 18617, GN 92 of 1998.

[136] See *Government Gazette* 18277, GN R1332 of 1997. While ICASA is currently in the process of finalizing this with respect to VANS providers, the new Ministerial Policy Directions have proposed that all telecommunications licensees, including VANS, shall contribute a percentage of turnover prescribed by regulation, but shall not exceed 0.5%. The current ICASA proposals for VANS requires 0.25%. See *Government Gazette* 21642, 11 October 2000.

The Universal Service provision in the RP however, is not without criticism: commentators have identified the vague language used to articulate this principle, as a potential barrier to market entry. For example, failure to specify the maximum set of elements comprising universal services; the amount of subsidy and how to determine who should be a universal service provider, allegedly allows Members to design measures that might serve as a disguised form of protectionism for local firms.[137] While the right of governments to determine domestic policy will be discussed below, I suggest that this criticism is lacking in foundation. It is precisely because the value of universal service across countries, with varying levels of development, is difficult to quantify, that this area of policy making must fall squarely within the domestic domain. The complexities inherent in designing an effective universal service policy are too context specific to be determined at a multilateral level. To do so, would seriously compromise the pursuit of these social and other non-economic objectives.

### 4. Public availability of licensing criteria

Like many other countries, SA requires that all telecommunications services be provided under a license issued in terms of the Act.[138] As licensing mechanisms can also serve as barriers to market access, the RP requires that the criteria, terms and conditions for licensing, where applicable, must be publicly available and where an application is denied, reasons must be provided on request. This requirement is valuable both from an administrative fairness point of view and to create certainty for investors. This is even more important in a market such as SA's where the realization of a domestic policy goal, such as ownership by historically disadvantaged persons, will remain a fixed criteria for awarding at least major service licenses.

In SA, the only two service categories that are currently open to full competition, subject to licensing, are VANS and PTNs. It is in respect of both these categories that the transparency requirements of the RP are being frustrated through delays in the finalization of a licensing regime.

---

[137] See Fredebeul-Krein and Freytag, at 631. See above note 91.

[138] Section 32(1). Section 33 specifies the categories of licenses which may be granted and the services authorized by such licenses. These are: PSTS; national and international long distance; local access and public pay phone services (all currently only applicable to Telkom) and cellular; VANS; and PTN services which may be provided on a competitive basis, but subject to licensing requirements.

At the time SA committed to liberalizing VANS services in its 1994 GATS undertakings, these services were authorized through individual agreements with Telkom. The initial list of sector specific commitments made[139] was later crystallized into the range of services that could be provided by a VANS licensee under the 1996 Act,[140] and although slightly modified, the list of services Telkom could provide under its VANS license.[141] In terms of the Act, ICASA is required to prescribe the form and manner of applications,[142] terms and conditions,[143] and where a license is denied or granted subject to conditions, the written reasons therefor.[144] Textually, SA's licensing requirements certainly approximate the guideline principles of the RP. To date however, a licensing framework for VANS and PTNs is still not yet publicly available. While the process was completed by ICASA in the second quarter of 2001, the regulations await approval by the Minister in order to be gazetted as law.[145] In the absence of a framework however, interim VANS licenses have been automatically granted to domestic and foreign suppliers, under interim SATRA guidelines.[146]

Thus, it is unlikely that the delays could be viewed as amounting to lack of compliance with the RP. There are however two important implications that flow from them: the first is that the uncertainty has largely contributed to the SAVA and other disputes presently under consideration. The second is that the lacuna has informed ICASA's current attempt to design a framework by addressing the nature of the VANS disputes that have emerged. Specifically,

---

[139] GATS, *Schedule of Specific Commitments (South Africa)*, GATS/SC/78 (15 April 1994) (94-1075).

[140] Section 40. Although the Act does not define VANS services, s 40(2) lists as services under a VANS: electronic data interchange (EDI); protocol conversion, e-mail and providing access to a data base or managed data network service.

[141] The Telkom VANS License defines VANS as "all those value added services provided by [Telkom] prior to the commencement of the Act, including the four services listed above and *in addition*: voice mail; store and forward fax; videoconferencing; telecommunication related publishing and advertising services, whether electronic or print; electronic information services, including Internet Service provision, and any other service in respect of which conveyance of signals is no more than is incidental to and necessary for, the provision of that service.

[142] Section 34(1).

[143] Section 34(3)(b).

[144] Section 35(6)(a). As with the RP, written reasons to be furnished only on request.

[145] *Government Gazette* 8462 of 1997, 'Notice in Respect of a Regulatory Framework for VANS' and *Government Gazette* 20866 of 4 February 2000, 'Notice of Intention to make VANS and PTN regulations in terms of section 96(2) read with section 34(1) of the Telecommunications Act No. 103 of 1996.' ICASA held public hearings on 6 and 7 February 2001, following written submissions on the second draft of VANS and PTN licensing regulations.

[146] These guidelines required information regarding: the applicant, contact details; the extent (%) of beneficial ownership by historically disadvantaged groups, women and disabled persons; the description of the service; geographical areas of provision; projection of the market size and target groups for the proposed value added network service(s); forecast of revenue for the first year; technical information including configuration and interfaces, description of equipment and proof of a type approval license.

ICASA has proposed conditioning the granting of a license on the usage to which the facilities obtained under the VANS will be put.[147] This cautious, but arguably excessive regulatory effort by ICASA may create a number of problems for future compliance with the GATS instruments.

One additional point however warrants brief attention: Bronckers and Larouche have cited as a weakness the fact that the RP makes no provision for the mutual recognition of licences, possibly requiring regional operators to obtain individual licences from each country in a region. They do note however that even in established regional blocs like the EU, such mutual recognition has proven elusive.[148] It is arguable that this is a strength rather than a weakness, certainly in offering a degree of support to national policy formulation under the GATS. The domestic policy imperatives vary from region to region and between countries within regions. To have mutual recognition in regard to licensing would deny members the right to impose criteria that serve their particular domestic policy goals, such as foreign ownership restrictions or requirements for ownership by members of historically disadvantaged groups. To the extent that GATS is intended to embody notions of fair treatment and non-discrimination, it is sufficient that these principles operate at a relatively high level. Indeed, to do otherwise would not level playing fields, but arguably tip them in favour of dominant international players with the most resources.[149]

---

[147] Above note 145.

[148] Above note 20 at 31. They note that the EU proposal on licensing, *Common Position 7/97* has only mandated the simultaneous and co-ordinated issues of licenses in the Union.

[149] In this regard, it is important to note a recent allegation made by the USTR that SA is violating Article XVI of the GATS in proposing that VANS licenses contain a 15 per cent minimum shareholding by historically disadvantaged persons. The proposal is contained in *Government Gazette* GG N 4041, 11 October 2000, No. 21642. USTR's argument suggests that the 15% per cent minimum requirement amounts to an unscheduled limitation on the participation of foreign capital in sectors which it has scheduled GATS market access commitments. Their complaint is that SA has scheduled commitments in the relevant sector, but has not listed any exceptions applicable to the measure in question. (Article XVI:2(f)). This claim is controversial and arguably inaccurate: at the time of scheduling the relevant sector, no formal policy on VANS existed and SA reserves the right to introduce such policy, which may in future introduce a limitation on shareholding. On the language employed by SA in the schedule, and the fact that Article XVI is sector specific, it is plausible that SA has in fact not made any substantive specific commitment. This argument is supported by the European Commission's "Info-Point" on World Trade in Services representation of country commitments. In their account of country commitments by sector, they indicate that SA made no commitment in the services that constitute those currently defined as VANS. Finally, while autonomous liberalization beyond that which SA specified may have occurred, this does not however, attract obligations under the GATS. The USTR also noted that the requirement is a disincentive to investment by foreign companies. This too is a contested claim. This matter has not as yet, been raised in a formal complaint and due to length constraints cannot be examined in any greater detail here.

*5. Independent Regulators*

It is clear from a consideration of the foregoing that an institutional, non-aligned arbiter is essential to the implementation and maintenance of a competitive market; to license, monitor and enforce obligations and conditions, and to create a forum for dispute resolution. The RP thus requires an independent regulatory body whose decisions and procedures will reflect this ideal. The fact that the RP, however, is silent on the need for regulators to be separate from and not have to be accountable to government, as well as to major suppliers has been raised as a criticism of its content.[150] Melody points out that while 'independence' has a number of facets, it does not necessarily include the power to make policy, "but rather the power to *implement* policy without undue interference from politicians, or industry lobbyists.[151] This accords with the approach to institutional design followed in SA. [152] Whilst the provisions of the RP in this regard remain vague, accepting this characterization of independence goes some way to alleviating these anxieties. Yet, even on this interpretation, SA's compliance in this regard requires some comment.

In 1993, prior to the first democratic elections in SA, an agreement brokered at the multiparty negotiations, saw the establishment of an independent broadcasting authority, (IBA)[153] whose primary role was to ensure free and fair broadcasting and media coverage of the upcoming elections. As a result of the sentiment at the time, the IBA was given the power to *formulate*, as well as implement broadcasting policy independently, but like SATRA was to receive its operational budget from Parliament.[154] In addition, section 192 of the SA Constitution[155] entrenched the status of an independent broadcasting regulator.[156]

The White Paper for Telecommunications in 1995 had contemplated a merger between the two distinct authorities, which was effected in July 2000.[157] However, the constitutional entrenchment in section 192 coupled with their historically different policy making roles, raised serious doubts over the continued independence of the IBA. Further political wrangling occurred over the provisions of the proposed Merger Bill,[158] playing to a range of perceptions regarding the third cell debacle and SATRA's overall financial management and regulatory ability.[159] Nonetheless, the importance of the merger was generally accepted as a prudent rationalization of resources in light of technological convergence. Uncertainty as to the IBA's independence was eventually resolved, at least textually, by including the words "as required by section 192 of the Constitution" in its main object of regulating broadcasting in the public interest.[160] Section 3 of the Merger Act offered further sop in stipulating that the new Authority function without political or commercial interference.[161]

---

[150] Bronckers and Larouche, at 31.
[151] Melody, above note 101 at 25.
[152] Telecommunications Act 1996, s 5(3).
[153] The Independent Broadcasting Authority Act, 1993.
[154] Section 3(3) states, that "The Authority shall function without any political or other bias or interference and shall be wholly independent and separate from the State, the government and its administration or any political party."
[155] The Constitution of the Republic of South Africa, 1996.
[156] This so-called Chapter 9 provision forms part of the section conferring rights on State institutions supporting Constitutional Democracy, including for example, the Public Protector and the Human Rights Commission.

The issue of funding however remains a serious obstacle. As were its predecessors, ICASA is wholly dependent on Parliament for funding.[162] Like its predecessors, ICASA has continued to cite the lack of financial resources as a serious impediment to effective regulation. International consultants on the Merger noted that [the agency's] statutory financial ties to the Department of Communications "does little for independence".[163] In absolute terms however, one is forced to question the extent to which a regulatory agency can ever be truly independent whilst on a government allowance? This is of course an issue that affects the full range of independent regulatory agencies and is not limited to telecommunications. Arguably the problem is not that public money is used but rather the attendant powers that inhere in the allocation of these funds and the implications this has for policy implementation.

---

[157] Independent Communications Authority of South Africa (ICASA) Act, 2000.

[158] Independent Communications Authority of South Africa Bill, 2000 [B 14B-2000]. Parliament tried, unsuccessfully however, to give the powers of appointment and removal from office to the executive. There was also debate over a clause in the Telecommunications Act that allows regulatory decisions to stand even if an improper interest is established on the part of a councilor. This too was scrapped in the final Act. See Barry Streek, 'New Communications Bill Amended' *Mail and Guardian,* 14 April 2000.

[159] Some assert that SATRA was considered "too" independent for government; others suggest that the merger was timeous in order to deal with allegations of impropriety and to "renew a quality in leadership to give confidence to a rather shaken industry". See 'All Eyes on the appointment of Independent Regulators' *World Reporter,* 11 May 2000.

[160] Independent Communications Authority of SA (ICASA) Act, 2000, s 2(a).

[161] Section 3(4). Section 3(3) suggests that ICASA be only subject to the Constitution and the law. However, see below note 164.

[162] Section 15(1) read with attached memorandum on the objects of the Independent Communications Authority of South Africa Bill, 2000 [B 14B-2000]. ICASA is required to operate off the combined budget allocations received by the IBA and SATRA as separate entities.

[163] Alan Darling, above note 84. See also Marina Bidoli, 'We're breaking up' *Financial Mail,* 23 April 1999.

The misfortunes of the third cellular process and the flurry of court cases in the VANS arena, have been edifying on a number of levels, but for the most part reflect the ease with which regulatory capture can occur, or be perceived to be occurring.[164] It is noted that the efficacy and survival of regulators depends on their success in earning legitimacy in the eyes of the public and the stakeholders. Without that legitimacy, independence as well as the ability to perform core functions is effectively threatened.[165] The extent to which the capture allegations against ICASA prove either true or false is relevant to SA's obligations under its WTO commitments, but lack of compliance must be located within the context of the agency's funding.[166] To safeguard independence, the RP could have contemplated the inclusion of a neutral funding mechanism, such as the revenue generated from the payment of license fees.

### 6. Allocation and use of scarce resources

The RP includes commitments to allocate and use frequencies, numbers and rights of way in an objective, timely, transparent and non-discriminatory manner. This directive can be linked to "global commons" theory, namely that effective management and allocation is necessary in order to co-ordinate a range of services and users who share a common (and scarce) resource. Inefficiencies in one jurisdiction can potentially spill over into others infringing on the overall effectiveness of the resource for all.[167]

ICASA is the agency responsible for control and management of the radio frequency spectrum and is obligated to honour all international agreements and standards in that regard.[168] To this end, the Authority has undertaken a Band Re-planning Exercise, Phase I of which was initiated under SATRA.[169]   Similarly, in 1997 SATRA began a review of SA's numbering policy with a view to the development of a new numbering plan, which despite its completion over a year ago, is still awaiting promulgation.[170] The finalization of this plan will have to occur prior to the licensing of the second network operator, as numbering policy, particularly in regard to portability and carrier pre-selection, is crucial to fixed line competition. The proposed Ministerial Policy Directions contain a number of provisions regarding numbering, including the introduction of number portability in April 2003.[171]

---

[164] Arguably as a direct result of the third cellular licensing debacle, the Schedule to the Telecommunications Amendment Bill contains a revised and expanded list of grounds upon which a councilor may, at any time, be disqualified or removed from office. It is worth noting however that this amendment authorizes the Minister, rather than the President, as is the case in the current Act, to effect the disqualification. The Minister merely requires the approval of the President and the concurrence of the National Assembly and Portfolio Committee.

[165] Rohan Samarajiva, 'Establishing the legitimacy of new Regulatory Agencies' 24 Telecommunications Policy at 183 (2000).

[166] It is worth noting that for the court review of the third cell process, ICASA had to approach Parliament to increase its budget in order to pay legal fees to defend its actions.

[167] See Trebilcock and Howse, above note 6 at 422.

[168] Telecommunications Act, s 28. Section 30 requires licensing for specified radio transmission and reception services.

## IV. CONFLICT AND IMPLEMENTATION

It becomes apparent through a review of SA's compliance with its GATS commitments that any apparent lag is attributable both to the specific conditions in SA and also to a number of general problems inherent in the RP. While a number of operational flaws have been highlighted, they can all be reduced to one common theme: the level of generality with which the Reference Paper is phrased. To the contrary, rather than view this as a shortcoming, this generality is essential to allow Members to pursue national development policies so crucial to their political legitimacy. This facilitates the development of the tension alluded to throughout the preceding analysis and is addressed below.

### A. Limitations of the Fourth Protocol

Whilst solutions may diverge, there is apparent consensus amongst critics of the Fourth Protocol and the RP as to its weaknesses.[172] These include the vagueness of the competition and interconnection principles; the voluntary nature of the specific commitments; the ability to file extended exemptions; the potential to free ride; to erect disguised and sanctioned barriers through numbering, licensing and universal service policies; the failure to specify regulatory independence from government and the opportunity to phase in changes and thus protect monopolies, all of which may frustrate the GATS in achieving its goals.[173]

---

[169] The second South African Band Re-planning Exercise (SABRE 2), *Government Gazette* 21833, GN 4568 of 7 December 2000.  SABRE 1 was completed in 1997.

[170] *Government Gazette* 20354, GN 1771 of 1999 and *Government Gazette* 19786, GN R198 of 1999. SATRA published a discussion paper on the future of numbering in SA and drafted a final plan at the end of 1999.

[171] ICASA regulations dealing with number portability, carrier pre-selection and facilities leasing were drafted for public comment in July 2001. See *Government Gazette* 22533 GN 1781 of 2001. ICASA however withdrew its public inquiry following the joint policy statement issued on 15 August 2001 regarding further changes to the proposed telecommunications policy. See *Government Gazette* 22588 GN 1890 of 2001.

[172] Specifically reference is made to Bronckers and Larouche; Fredebeul-Krein and Freytag, and in defense of the RP, Blouin, above note 99.

Noam and Drake critique the nature, rather than the form of these commitments as "standstill" consisting of binding liberalizing measures that have already been adopted at the national or regional level. [174] Whilst this is true for many countries, the Schedule however, does at least create a fixed (and arguably stricter) timetable for their realization. Nonetheless, the sum total of the aggregated weaknesses is two-fold: a clear doubt as to the effectiveness of a general commitment to foster and promote competition in telecommunications, and secondly, that the principles on their own, without sufficient particularity, could have the adverse effect of promoting protectionism endorsed, albeit unwittingly, by the GATS regime.

It is difficult to deny purchase in this view. However, the point that these criticisms appear to gloss over is precisely the fact that the document is a *guideline*. It is explicitly due to the asymmetry inherent in the WTO treaties that a flexible guide is needed to manage and allow for these differences while establishing a minimum common standard. It is for this reason that the proponents of the view that the RP should be more specific are, I submit, wrong. To give effect to these ideals, like the provision of dial tone in remote rural areas of South Africa, requires flexibility and pliancy in policy formulation. It is also for this reason that the nature of 'independence' may vary in its degrees of separation from government, as long as it is at arm's length;[175] that the structure of universal service objectives and who is liable to effect them, should be decided at a national level and that the licensing criteria for the provision of service should be capable of local tailoring. As long as a minimum floor of agreed standards is place, which accord with principles of administrative fairness, domestic governments should have the leeway to craft policies that best give expression to broader macroeconomic and social goals. For these reasons, it must be accepted that the RP was included in the negotiations to provide just the necessary *safeguards* in domestic law, not to override or supplant it.[176]

---

[173] See Gates, above note 12 at 99 and Senunas, above note 6. Slightly expanded, these critiques suggest that the safeguard provisions lack definition as to what constitutes anti-competitive behaviour, when cross-subsidization is impermissible or how such behaviour should be addressed. Similarly the lack of criteria for interconnection is so vague as to render cost-based pricing difficult, if not impossible to realize in practice. The further lack of specificity for numbering policy and licensing conditions erect potential barriers to entry. They also argue that the provisions of GATS Article VI, do little to put the criteria for licenses in concrete terms and they criticize the RP's silence on the enforcement powers and structural separation between the Ministry and the regulator. See Bronckers and Larouche, at 27-30; Fredebeul-Krein and Freytag at 628-629 and Sherman, above note 88 at 77. But see above note 104.

[174] E. Noam and W. Drake, 'Assessing the WTO Agreement on Basic Telecommunications' in G. Hufbauer and E Wada (eds) *Unfinished Business: Telecommunications after the Uruguay Round* (Washington; Institute for International Economics, 1997) cited in Chantal Blouin, above note 99 at 138.

Along these lines, other commentators are more sanguine. Blouin argues that the RP's role is merely one of an 'insurance policy' in that the commitment to keep a sector open and subject that undertaking to multilateral dispute settlement, provides investors with certainty and predictability that domestic liberalization alone cannot offer.[177] Whilst acknowledging that a horizontal approach to market openness may allow for a more coherent perspective on the relationship between trade, competition and regulatory policy, she concludes that the sectoral approach offers many more benefits. The most important of these is the recognition that including the specifics of competition policy and domestic regulation in multilateral trade agreements, would be viewed as an unacceptable encroachment on national sovereignty, which would in turn serve to undermine the very legitimacy of the international trade regime.[178] It remains a key issue of concern however, as to how effectively the WTO dispute settlement process is likely to interpret and apply open-ended principles such as those contained in the RP, both in terms of technical competency and political legitimacy.

**B. Benefits of the Fourth Protocol: Domestic versus International**

Without traversing this discussion, it is trite to observe that trade in services requires special consideration apart from trade in goods. Further, cross border projections of capital assets and individuals do raise certain political concerns, which in turn engenders a degree of government regulation.[179] I argue however, that recognizing the distinctiveness of trade in services at the intersection between international trade and domestic policy produces an understanding of why an inherent tension between the two emerges. Moreover, it is submitted that this has long been an obvious point to the multilateral trade regime. Indeed, textual support for its management is implied in the GATS: Article II specifically authorizes measures inconsistent with MFN subject to certain conditions; Article XIX explicitly directs that negotiations on specific commitments "shall take place with due respect for national policy objectives," which will include "flexibility" for opening fewer sectors, liberalizing fewer types of

---

[175] Rohan Samarajiva. Above note 165 at 183. See also W.H. Melody, above note 101.
[176] Janisch, at 72.
[177] Blouin, at 139.
[178] Ibid at 140.
[179] See above note 6.

transactions…and, attaching access conditions to foreign suppliers services.[180] The Annex also recognizes this tension within the developing country context and authorizes reasonable conditions to be placed on access to and use of those Members public telecommunications networks and services, "*necessary to strengthen* [their] domestic infrastructure and *increase* [their] *participation* in international trade in telecommunications services."[181] The new guidelines for subsequent services negotiations unequivocally endorse these fundamental GATS principles.[182]

In the SA context, analysis of this flexibility is apposite. On the one hand, macro economic policy commits the country to liberalization and privatization. This inevitably requires an environment attractive to foreign capital and investment. On the other hand, because of inequalities resulting from Apartheid, a "redress initiative" has been injected into almost all economic sectors. Telecommunications is no exception. The Act, dedicated to public interest regulation, lists a number of objects in this regard and requires their promotion.[183] While it is uncommon for legislation to embody such extensive enunciation of aims and objectives, it is important to understand the enormous financial stakes at issue and the influence of the historical context of Apartheid from which this emerged.[184] For example, until 1997, not a single spectrum license was owned by a black person or by a woman in SA.[185]

Measures to promote the realization of many of the Act's objectives, for example, universal service and ownership by historically disadvantaged individuals must inherently clash with the fundamental underpinnings of the GATS – market access and national treatment.[186] Universal service aims, as we have seen in the SA context, may serve to buttress a monopoly and contribute to sustaining a closed market.[187] Ownership restrictions similarly limit foreign entry. Yet, empowerment as a value must be a criterion used, alongside technical compliance and financial competence, to determine the basis upon which licenses should be awarded. This cannot be effected if the undiluted ideal of the GATS – open markets - were to be implemented without restraint. It is clear however, that this is indeed not the case and such domestic goals are thus not GATS incompatible: GATS explicitly provides for the expression of these ideals in order to create critical mass and support the political legitimacy of those signatories domestically.

---

[180] GATS, Article IV.
[181] Section 5(g). [Emphasis Added]
[182] The new Guidelines and Procedures for the Negotiations on Trade in Services were adopted by the Special Session of the Council for Trade in Services on 28 March 2001. See WTO Press Release, 217, 2 April 2001.
[183] See note 35.
[184] In 1997, world telecommunications revenue was estimated at 644 billion USD and global investment in telecommunications totaled 170 billion USD, of which Africa accounted for approximately 10 billion USD. See Council for Trade in Services, *Telecommunications Services* S/C/W/74, 8 December 1998. According to the ITU, in 1998, telecommunications revenue in South Africa was 5,971 million USD, accounting for 3.4% of GDP. In SA, telecommunications, broadcasting and information technology combined comprise a R58 billion industry and contribute increasingly to GDP, which in 1998 was R730 billion. Of this the telecommunications industry accounts for 39%, the IT industry 42% and the equipment supply side 19%. Figures derived from BMI-Techknowledge 1999 and 2000 South African Reserve Bank and cited by Alison Gillwald, "Building Castells in the Ether" [forthcoming in 2001].
[185] From what can be ascertained from the records. See Alison Gillwald, 'Telecommunication Policy and Regulation for Women and Development', Development Summit, TELCOM 99 + INTERACTIVE 99, Geneva, October 10- 17, 1999.

In this light, for SA to have been denied the flexibility to prescribe the content of the regulatory principles it undertook, with due regard to transparency and competitive neutrality, would have completely undermined its domestic reform agenda as well as the expressed recognition of these ideals of the GATS, cited above. At the same time, the aim of competition and attracting investment cannot similarly be realized by application of domestic policy alone – a policy that explicitly endorses the dominance effects of a monopoly market. Thus, to not have the RP's principles with a sufficient degree of particularity to establish regulatory standards necessary for international competition, would have been equally defeatist of the GATS purpose. What emerges from that tension is a useful mechanism to attempt to balance domestic objectives with international trade commitments. This is particularly useful in a context that requires sensitivity to systemic inequality and the foresight to prepare for international competition. Many may argue however, that this tool has not been successfully applied, particularly in light of the foregoing analysis. This does not however, detract from the argument that the tension itself can be an important facilitating tool for domestic development, in the SA instance for example, to check the 'cream skimming' effects of full deregulation in a developing market and at the same time gradually to open it to competition.

---

[186] Notwithstanding the debate regarding bindings, it is within this context that the USTR's allegation pertaining to the 15 percent empowerment shareholding limitation, must be viewed, above note 149. Given the importance of the empowerment component in licenses, and the role of the 'redress initiative', it is arguable that in the unlikely event that a panel were to find SA bound by its 1994 offering, SA should request consultations and apply for a modification to its schedule, subject to a compensatory adjustment. While the wisdom of this may be questioned from a commercial and competitive angle, the ability to draft national policy giving effect to this goal, must be secured.

[187] While the monopoly, cross-subsidization model has been seen as a common way to finance universal service, it is not the only way to achieve this goal. In the SA context however, this has been both the approach and the justification for alleged and perceived anti-competitive behavior by Telkom since 1996.

Less than satisfactory results in SA's case, I argue are attributable to weak regulatory design and implementation, but this is not a burden the regulator should bear alone: the legislature needs to loosen the purse strings and clothe the regulator with ability to carry out its functions. The executive and the regulator need to find acceptable ways to accommodate the blur between policy formulation and implementation, without creating interference in each other's domain. It is suggested, though, that the lessons emerging from the first three years typify those that occur in newly liberalizing sectors adjusting to the demands and complexities of re-regulation. It is hoped that these will diminish over time as this adjustment is gradually achieved.[188]

## V. CONCLUSION

The last decade has seen more policy and legislative reform in international telecommunications than any other time in history. The GATS Fourth Protocol is a clear reflection of how extensive and far-reaching this "reformation era" in telecommunications is. I have argued, however, that this global framework, predicated on opening markets and preparedness for competition, presents substantial difficulties for countries whose development imperatives manifest a seeming conflict with this ideal. I have endeavoured to illustrate this dynamic by assessing the degree to which SA has complied with its GATS undertakings and have tried to explain any apparent lag in compliance by illustrating this tension at play.

To date, there are clearly areas where SA has ostensibly skirted non-compliance.  There are also potential risks for non-compliance with other commitments in the future, but for these, it is too early in the liberalization process to make a definitive assessment. I have also argued that SA never undertook any commitments more onerous than those it already assumed through its process of telecoms reform from 1994 until the present, although I have argued that GATS imposes a more rigorous timetable for implementation of such reform.

---

[188] The ITU has initiated a Regulator's Forum for existing regulators and policy-makers interested in establishing a regulatory body. The first Development Symposium for Regulators was held in Geneva from 20-22 November 2000,

Further, I have endeavored to show that the dynamic produced by the tension between domestic policy and international trade objectives is a useful mechanism to effect reform on the domestic level. It is in this context that the tension is asserted to be a positive one, operating as an effective check and balance against the deleterious effects that either, left to their own devices, can create. However, from the above assessment of SA's GATS commitments, it is clear that greater political will and policy co-ordination will have to be displayed, to garner the benefits of this application. Under this line of argument, it is apparent that similar domestic tensions, evidenced by the range of disputes before ICASA and the courts, exacerbated by a weak regulatory agency, is less beneficial, and potentially damaging to the sector. It is imperative that SA "get the regulatory framework right" and re-establish the legitimacy of an independent authority, so crucial to successfully managing and maintaining momentum in the transition to competitive markets.[189]

In terms of a comparative assessment of SA's compliance, the WTO Trade Policy Review Mechanism should be instructive when its next cycle occurs. Aimed at improved adherence to rules, disciplines and commitments made under multilateral trade agreements, this body assesses individual members trade policies and practices and how they have an impact on the trade regime. SA's review cycle is set to occur every four years and was last reviewed in 1998 as part of a grouped review of the Southern African Customs Unions (including Botswana, Lesotho, Namibia and Swaziland).[190] Arguably, this review was held too soon after undertakings were made to be of significant value. The next review however is scheduled for 2002 and corresponds with the second phase of SA's telecoms liberalization - the expiry of Telkom's monopoly and the limited introduction of fixed line competition.

There are also, however, two important factors to consider in assessing compliance at such an early stage of the Schedule's operation. The first is that the Fourth Protocol itself is embryonic and the augmented GATS agreement is inchoate in many respects. Its rules are not complete; they remain largely untested and consequently, there is no informative body of

---

the goal of which was to launch a dialogue in which all the world's regulators can share their experiences and views in order to learn from each other. See http://www.itu.int/treg. (visited 25 November 2000).

[189] See the ITU, *Trends in Telecommunication Reform: Convergence and Regulation* (1999).

[190] WTO, Press/TPRB/74. This took place between 21-23 April 1998. An account of the Chairperson's conclusions is available at http://www.wto.org/english/tratop_e/tpr_e/tpr_e/tp74_e.htm. (visited 1 December 2000).

precedent or jurisprudence in telecommunications. Subsequent service negotiations in telecommunications have been delayed and are relatively new in the overall multilateral framework.[191] With experience, many gaps in the instruments will be identified and will need to be addressed and in some cases, improved. Certainly, several more rounds of negotiations will be required. Secondly, whatever the perceived weaknesses of the GATS Fourth Protocol may be, because the agreement was negotiated as part of a multilateral treaty, offers and commitments are binding and practically irreversible. In this light, it reflects a minimum consensus on the goals to be pursued and the lowest common denominator below which a signatory cannot go. Because of the critical importance of telecommunications as an essential infrastructure to the information economy, this consensus, embodied in a multilateral agreement, has an inherent and momentous value of its own.

---

[191] WTO Members are currently tabling proposals regarding both the structure and the contents of the new 2001 negotiations. By the end of the first quarter, 70 proposals had been received from 40 countries. See http://www.wto.org/english/tratop_e/serv_e/s_propnewnegs_e.htm.